\documentclass[lettersize,journal]{IEEEtran}
\pdfoutput=1
\usepackage{amsmath,amsfonts}
\usepackage{algorithmic}
\usepackage{algorithm}
\usepackage{array}
\usepackage[caption=false,font=normalsize,labelfont=sf,textfont=sf]{subfig}
\usepackage{textcomp}
\usepackage{stfloats}
\usepackage{url}
\usepackage{verbatim}
\usepackage{graphicx}
\usepackage{cite}
\usepackage{hyperref}
\hypersetup{hidelinks,
	colorlinks=true,
	allcolors=black,
	pdfstartview=Fit,
	breaklinks=true}
\usepackage{stfloats}
\usepackage{amsmath}
\usepackage{booktabs}
\usepackage{comment}
\usepackage{ragged2e}
\usepackage[caption=false, font=normalsize, labelfont=sf, textfont=sf]{subfig}
\UseRawInputEncoding
\begin{document}

\title{Joint Optimization of Base Station Clustering and Service Caching in User-Centric MEC}

\author{Langtian Qin, Hancheng Lu,~\IEEEmembership{Senior Member,~IEEE}, Yao Lu, Chenwu Zhang and Feng Wu,~\IEEEmembership{Fellow,~IEEE} 
\IEEEcompsocitemizethanks{\IEEEcompsocthanksitem L.Qin, H.Lu, Y.Lu, C.Zhang and F.Wu are with the Department of Electronic Engineering and Information Science, University of Science and Technology of China, Hefei 230027, China. (E-mail: qlt315@mail.ustc.edu.cn; hclu@ustc.edu.cn; luyao98@mail.ustc.edu.cn; cwzhang@mail.ustc.edu.cn; fengwu@ustc.edu.cn)  \protect\\

 }
}

\maketitle
\begin{abstract}
Edge service caching can effectively reduce the delay or bandwidth overhead for acquiring and initializing applications. To address single-base station (BS) transmission limitation and serious edge effect in traditional cellular-based edge service caching networks, in this paper, we proposed a novel user-centric edge service caching framework where each user is jointly provided with edge caching and wireless transmission services by a specific BS cluster instead of a single BS. To minimize the long-term average delay under the constraint of the caching cost, a mixed integer non-linear programming (MINLP) problem is formulated by jointly optimizing the BS clustering and service caching decisions. To tackle the problem, we propose JO-CDSD, an efficiently joint optimization algorithm based on Lyapunov optimization and generalized benders decomposition (GBD). In particular, the long-term optimization problem can be transformed into a primal problem and a master problem in each time slot that is much simpler to solve. The near-optimal clustering and caching strategy can be obtained through solving the primal and master problem alternately. Extensive simulations show that the proposed joint optimization algorithm outperforms other algorithms and can effectively reduce the long-term delay by at most $93.75\%$ and caching cost by at most $53.12\%$.
\end{abstract}


\maketitle

\section{Introduction}
\IEEEPARstart{W}{ITH} the rapid development of the mobile Internet, data traffic is experiencing explosive growth due to pervasive mobile devices, ubiquitous social networking, and resource-intensive applications\cite{Yao2019}.  When running newly emerging applications such as augmented reality (AR) and virtual reality (VR), massive computing tasks will be generated (video rendering, etc.). Processing some of the computing tasks depends on various types of services. For example, in an AR application, the object databases and visual recognition models are required to process the user’s input data and run classification or object recognition\cite{Zhang2021}. However, in rush hours or traffic jams, direct service dissemination from remote data centers in real-time may lead to unprecedented network traffic load and congestion, and may also induce long transmission delay\cite{Wu2022}. By deploying servers in the radio access network (RAN), mobile edge computing (MEC)\cite{luo2021survey,shi2016survey} can provide users with low-latency computing, caching, and transmission capabilities\cite{hu2015white}. MEC servers can pre-cache the popular services in advance, and process user’s offloaded requests instead of routing the requests to the remote data centers\cite{samanta2019adaptive}.  Through caching services in a distributed manner that is close to users, edge service caching overcomes the problems of high transmission delay caused by long-distance data transmission of the cloud server, alleviates the burden on the backhaul links, and reduces the risk of being attacked by the malicious nodes.

Existing works\cite{samanta2019adaptive,nath2020deep,xu2018joint, Xuanheng2022,Qi2019,Yang2022,bi2020joint,castellano2019distributed,meng2019dedas,huynh2021joint,Pou2019,Zhu2019} on edge service caching have mostly adopted the cellular-based edge caching networks where multiple users are served by a single MEC-enabled base station (BS) in the cell. However, the performance of edge service caching is greatly limited by the wireless transmission in traditional cellular-based networks. Firstly, the cached services can only be provided by the BS that the user accesses. Since service caching needs to consume computing and storage resources, resource-limited BS can only cache a small set of services at the same time, which is prone to being unable to cache the services required by users. Although the services can be transmitted or relayed through other BSs, it will consume additional transmission overhead. Secondly, users at the edge of the cell will suffer severe signal attenuation and inter-cell interference, which leads to a reduction in the data transmission rate. The large signal interference may even result in transmission failure, which will affect the quality of service (QoS) of users. Although some works have proposed cooperative edge service caching\cite{Luo2022,Li2022,Xiao2022,Lin2022,Gao2022}, they still ignore the impact of wireless transmission in cellular networks on the system.

As a key technology in 5G and beyond\cite{Chen2014}, user-centric network (UCN)\cite{chen2016survey,Pan2018,Su2015} breaks the concept of ``cell'' in traditional cellular-based network, which can be seen as a reliable solution to the above problems. In UCN, each user will be served by a dynamically divided BS set, which is called BS cluster\cite{papadogiannis2008dynamic,lin2014dynamic}. Each BS cluster can be divided adaptively according to the location and network condition of users to provide seamless wireless transmission service. By integrating the MEC server in the BS, UCN can further expand the computing and caching resource for task offloading and service caching in MEC. Compared with a single BS, BS cluster can store more types of services, thus can improve the successful probability of task offloading at the edge and save the corresponding caching cost. Moreover, BS cluster can jointly decode the signals transmitted by users, which can effectively offset the impact of intra-cell interference and ensure the efficiency of wireless transmission. Therefore, the user-centric edge service caching can provide users with efficient and reliable wireless transmission, service caching and task processing wherever they are.

 However, to maximize the system performance, the user-centric edge service caching still faces some challenges. Firstly, the caching strategy needs to be considered from a long-term perspective to meet the personal preferences of most users as much as possible. However, the channel state information (CSI) and the type of services requested by users are time-varying in the long-term decision-making process, and yet the decisions have to be made without foreseeing the future system dynamics\cite{Huang2012} \cite{Liu2016}. Secondly, BS clustering needs to be dynamically adjusted according to the changing network state information. In addition, for edge service caching, BS clustering will be coupled with the service caching decision. Therefore, it is challenging to make the optimal BS clustering decision when facing so many influencing factors. Last but not the least, after the BS cluster of the user is determined, the BS cluster needs to jointly provide transmission and caching services for the user. Different MEC nodes are heterogeneous in geography and resource capability, which also makes it difficult to design the cooperation mechanism in UCN.\cite{Tan2019,Tanzil2016}.

 To address aforementioned challenges, we proposed a novel user-centric edge service caching framework where each user is served by a specific BS cluster cooperatively. To minimize the long-term average delay under the constraint of the caching cost, we jointly optimize the BS clustering and service caching decisions. The main contributions of this paper are summarized as follows:

\begin{itemize}
    \item A user-centric cloud-edge cooperative task offloading and service caching framework is proposed to provide users with effective and reliable wireless transmission and service caching. We analyze the long-term delay and service caching cost starting from the single-user scenario to the multi-user scenario. The joint optimization problem is formalized to minimize the long-term offloading delay under the constraint of caching cost by optimizing the BS clustering and service caching decisions.
    
    \item To solve the problem, we transform the long-term joint optimization problem into multiple instantaneous problems using Lyapunov optimization. To implement the clustering and service caching in each time slot, we decompose each instantaneous problem into two sub-problems (i.e., the primal problem and the master problem) based on generalized benders decomposition (GBD) and design an efficient algorithm called JO-CDSD to solve the sub-problems alternately. JO-CDSD can obtain the near-optimal strategy without future information.
    
    \item We conduct extensive numerical simulations to verify the effectiveness of the proposed algorithm both in the single-user and multi-user scenarios. Simulation results show that the proposed algorithm outperforms reference algorithms and can significantly reduce the long-term offloading delay and caching cost.
\end{itemize}

The rest of this paper is organized as follows. Section II reviews some related works. Section III gives the system model and the problem formulation. A joint BS clustering and service caching optimization framework is proposed in Section IV. Section V presents the evaluation results and analysis, followed by the conclusion in Section VI.

\textit{Notations}: We use $\boldsymbol A$ to represent a matrix. $a_{ij}$ denotes the element in $i$-th row and $j$-th column of $\boldsymbol A$, $\boldsymbol A^{\dagger}$ and $\Vert \boldsymbol A \Vert_2$ represents the Pseudo inverse and $\ell_2$ norm of $\boldsymbol A$, respectively. Specifically, we use $\boldsymbol I_N$ to denote a identity matrix with dimension $N$. We use $\boldsymbol x$ to denote a vector, and $x_i$ denotes the $i$-th element of vector $\boldsymbol x$. $\boldsymbol x^{T}$ represents the transpose of vector $\boldsymbol x$. $\mathbb{R}^{M \times N}$ and $\mathbb{C}^{M \times N}$ represents the space of $M \times N$ real and complex number matrices. We use calligraphy upper-case letter such as $\mathcal{M}$ to represent a set.

\section{Related Work}
Service caching in MEC has been extensively studied by lots of previous work\cite{samanta2019adaptive,nath2020deep,xu2018joint, Xuanheng2022,Qi2019,Yang2022,Xia2021,Xu2022,Xu2022Near}. In \cite{nath2020deep}, a MEC caching assisted offloading scenario is considered, where edge nodes can cache the tasks according to the popularity to avoid repeated caching. To minimize the weighted sum of energy consumption, delay, and cost, the author proposed a distributed algorithm based on Deep Deterministic Policy Gradient (DDPG) for decision-making. The authors in \cite{Yang2022} proposed a blockchain incentive scheme and introduced the  Stackelberg game to optimize the benefits for both the edge computing server (ECS) and D2D users in the blockchain network. In \cite{xu2018joint}, the authors discussed the challenges faced by service caching in densely deployed cells, including service heterogeneity, unknown system dynamics, spatial demand coupling, and distributed coordination. To solve the above problems, the authors jointly optimize service caching and offloading strategies. service offloading in a highly dynamic vehicle network is considered in \cite{Qi2019}, where multiple heterogeneous vehicle nodes act as MEC nodes, and the task offloading decisions is optimized by a knowledge-driven deep reinforcement learning algorithm. In \cite{Xia2021}, the authors propose Online MEDC (OL-MEDC), an approach that formulates Mobile Edge Data Caching (MEDC) strategies from the app vendor's perspective. The authors in \cite{Xu2022} study the service caching problem with request rate uncertainty, and propose an approximation algorithm and a Stackelberg game via leveraging the randomized rounding technique. To minimize the social cost of all network service providers, the authors in \cite{Xu2022Near} devise a distributed and stable game-theoretical mechanism for the problem with Virtual Machine (VM) sharing among network service providers.

The solutions\cite{bi2020joint,castellano2019distributed,meng2019dedas,huynh2021joint,Pou2019,Zhu2019} focus on the wireless transmission when optimizing the service caching strategy. The authors in \cite{huynh2021joint} optimized the task offloading, data content caching, computing resources, and transmission power allocation jointly in a MEC-enabled non-orthogonal multiple access (NOMA) network. In \cite{bi2020joint}, the authors considered the service caching problem in a single edge server with limited resources. The author formulated the service deployment problem into a mixed-integer nonlinear programming problem and jointly optimizes the offloading, service caching, and resource allocation decisions. In \cite{Pou2019}, the author jointly optimized the service caching and the request routing of user offloading tasks. In particular, the wireless transmission of offloading requests is considered when the ranges of base stations are independent or partially overlapped.

With the dense deployment of BS, a few works also discussed clustering or cooperative service caching in the ultra-dense network. By taking into account the cooperation among small-cell base stations (SBSs), the authors in \cite{Luo2022} proposed that an active SBS should share the observed rewards to the SBSs while an inactive SBS can keep learning the information of its surrounding users, and will become active again when its surrounding users have sufficient requests. A hierarchical hybrid transmission scheme is proposed in \cite{Li2022},  where users can obtain the requested layer files through the cooperative cellular multicast-D2D and THz transmissions. In \cite{Xiao2022}, the authors proposed a novel transcoding-enabled VR video caching and delivery framework for edge-enhanced NOMA-based wireless networks, and adopted multi-agent reinforcement learning to obtain the optimal caching strategy. The authors in \cite{Lin2022} proposed a novel cache-enabled user association scheme for hybrid HCNets with limited local storage, in which partial sub-6G local storage caches contents with relatively low popularity for higher overall content diversity. In \cite{Gao2022}, the authors proposed a heterogeneous multi-agent deep deterministic policy gradient (MADDPG) approach to optimize the cache storage and contents fetching strategy, which takes users and cache servers as two different types of agents to learn the cooperation and competition for mutual benefits.

Previous work based on cooperative edge caching was still unable to avoid the impact of inter-cell interference on wireless transmission in traditional cellular-based network. The optimization goals of the most existing works were always instantaneous delay or overhead, which is prone to ignore the needs of some users and the dynamics of the system. Different from the existing works, we design a collaborative edge service caching framework based on UCN and jointly optimize the BS clustering and service caching decisions. In addition, we focus on the long-term edge service caching and aim to minimize the long-term delay under the service caching cost constraint.

\section{System model and problem folmulation}

\subsection{Network and Services Model}
As shown in Fig.1, we consider a MEC-enabled user-centric wireless access network. We assume there are $M$ BSs and $U$ users, indexed by sets $\mathcal M$ and $\mathcal U$, respectively. Each user is equipped with a single antenna and each BS is equipped with $A$ antennas and endowed with MEC functionality. Therefore, BSs can provide certain computing and caching resources to users. Considering the heterogeneity of MEC nodes, each node has different computing and caching resources. The computing and caching resources of BS $m$ are denoted as $C_m$ and $S_m$, respectively. Consider the time-varying of the communication system, the long-term process $T$ is divided into multiple discrete slots $t$ ($1\leq t\leq T,t\in \mathbf Z$). The BS cluster of user $u$ is denoted as $\Phi_u(t)$ and the set of users served by $\Phi_u(t)$ is denoted by $\Omega_u(t)$. In this paper, coordinated scheduling/beamforming (CS/CB)  mode\cite{Ammar2022} is adopted to enable cooperative wireless transmission for users. For a specific user $u$, all the BSs in its cluster will receive the offloaded data from the user and decode it jointly by exchanging channel state information (CSI) through the backbone network. It should be noticed that a BS can provide services to different users at the same time, and the BS clusters of different users can intersect. In addition to user $u$, other users served by BS clusters $\Phi_u(t)$ are called intra-cluster users, i.e, the users belonging to $\{v: \forall v \neq u, v \in \Omega_u(t)\}$. Users other than user $u$ and intra-cluster users are called inter-cluster users\cite{zhu2018stochastic}.

We assume there are $K$ types of services stored in the remote cloud, and the set of services is denoted as $\mathcal K$. Different services will consume different caching resources and computing resources when cached on the BSs. Let the caching and computing resources required by service $k$ as $\{s_k,f_k\}$. We use a continuous variable $x_{k,m}(t)\in [0,1]$ to express the probability that service $k$ will be cached on BS $m$ at slot $t$, and the service caching strategy of BS $m$ at slot $t$ is $\boldsymbol x_m(t)=[x_{1,m}(t),...,x_{K,m}(t)]^T$. 

\subsection{Task Offloading and Service Caching Model}
We assume that at time slot $t$, each user generates an offloading task $T_u (t)$ that requires only one type of service, and remain unchanged at each time slot. The data size and the workload of task $T_u(t)$ is denoted as $d_ {T_u(t)}$ (bit) and $w_{T_u (t)}$ (CPU cycles in GHz), respectively. Each BS needs to cache services at the beginning of each slot to satisfied the service requirement of the tasks. At each time slot, the entire offloading process of each user will go through three steps: task offloading, task processing, and result returning. Since the data size of processing result 
is relatively small, we ignore the downlink data transmission dalay in this paper. Therefore, the offloading delay consists of the following two parts.

\begin{figure}[t]
	\setlength{\belowcaptionskip}{-0.4cm}
	\centering
	\includegraphics[width=0.4\textwidth]{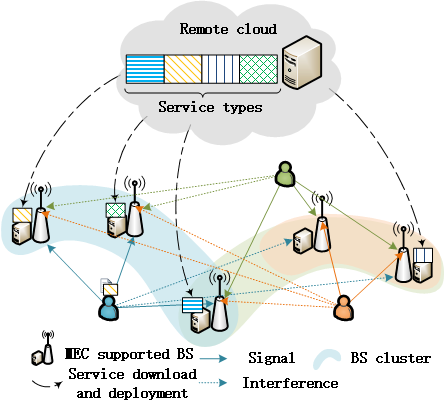}
	\caption{System illustration of user-centric MEC }
	\label{multi}
\end{figure}

\subsubsection{Uplink delay}
For BS cluster $\Phi_u(t)$, we use a binary variable $c_{u,m}(t)$ to represent BS $m$ $(m \in \Phi_u(t))$ serves user $u$ at time slot $t$ (1) or not (0), i.e., whether it belongs to the BS cluster of target user $u$ at time slot $t$.  Thus the clustering strategy of BS $m$ at time slot $t$ is $\boldsymbol c_m(t)=[c_{1,m}(t),...,x_{U,m}(t)]^T$. The total signal received by BS $m$ at $t$ consists of signals sent by target user $u$, intra-cluster users and inter-cluster users, which can be expressed as

\begin{equation}\label{signal}
	\begin{aligned}
		b_m(t)=&\sqrt{p_u}\boldsymbol g_{mu}(t)a_u(t)+\!\!\!\!\!
		\sum_{\tiny{\begin{array}{c}
					v\!\neq \!u,\\
					v\!\in \!\Omega_u(t)\end{array}}}\!\!\!\!\sqrt{p_v}\boldsymbol g_{mv}(t)a_v(t)\\
		&+\!\!\!\sum_{\tiny{\begin{array}{c}
					w\!\neq \!u,\\
					w\!\notin \!\Omega_u(t)\end{array}}}\!\!\!\!\sqrt{p_w}\boldsymbol g_{mw}(t)a_w(t)+n(t),
	\end{aligned}
\end{equation}

where $p_u$ is the signal power of user $u$, $\boldsymbol g_{mu}(t)\in \mathbf C^{A\times 1}$ is the channel coefficient between user $u$ and BS $m$, $a_u(t)$ is the symbol sent by user $u$, $n(t)$ is additive white Gaussian noise with 0 mean and variance $\sigma_n^2$.

Since all BSs in the BS cluster will share the CSI of the target user, the intra-cluster interference can be liminated by designing the coordinated beamforming vector. According to \cite{zhu2018stochastic}, the projection transformation zero-forcing beamformer of user $u$ is calculated as follows

\begin{equation}
	{\boldsymbol w}_u(t)=\frac{({\boldsymbol I}_{A\left|\Phi_u(t)\right|}-{\boldsymbol G}_{-u}(t)\boldsymbol G_{-u}^\dagger(t))\boldsymbol g_u(t)}{{\left\|({\boldsymbol I}_{A\left|\Phi_u(t)\right|}-{\boldsymbol G}_{-u}(t)\boldsymbol G_{-u}^\dagger(t))\boldsymbol g_u(t)\right\|}_2}.
\end{equation}

where $ G_{-u}=[\dots,\boldsymbol g_v(t)^ t,\dots]^T_{v\neq u,v\in \Omega_u}$ is the channel coefficient matrix between the BS cluster of intra-cluster users and the target user, and $\boldsymbol g_v(t)=[\dots,\mathbf g_{mv}(t),\dots]^T_{m\in \Phi_u(t)}$. The transmission rate of the target user $u$ can be expressed as

\begin{equation}
	\begin{aligned}
		r_u(t)=W log_2\Big(1+\!\frac{p_u\left|\boldsymbol{w}_u(t)^H\boldsymbol{g}_u(t)\right|^2}{\tiny \sum\limits_{w\notin \Omega_u(t)}
			\!\!p_w\!\left|\boldsymbol{w}_u(t)^H\boldsymbol{g}_w(t)\right|^2\!\!+\!\left|\boldsymbol w_u(t)\right|^2\!\sigma_n^2}\Big),
	\end{aligned}
\end{equation}
 where $W$ is the system bandwidth. Therefore, the uplink transmission delay of user $u$ can be written as

\begin{equation}
	D^{UCN}_{T_u(t)}=\frac{d_{T_u(t)}}{r_u(t)}.
\end{equation}

\subsubsection{Task processing delay}
Cache services on the MEC server requires BSs to pay a certain fee to the service provider. We assume the caching cost of service $k$ is proportional to the data size of service $s_k$. At time slot $t$, the total caching cost of BS $m$ is
\begin{equation}
    Cost_m(t) = \sum_{k\in \mathcal{K}} {\xi_k s_k x_{k,m}(t)},
\end{equation}
where $\xi_k$ is the caching cost coefficient. Assume that the offloading task is undecomposable, the task offloaded by each user will be processed by the BS with the highest probability of deploying such service in the BS cluster, i.e., BS $\hat m= \arg\max_m\{c_{u,m}(t)x_{\hat k,m}(t)\}$, where $\hat k$ represents the type of service requested at time slot $t$. BS has the probability of $1-x_ {\hat k,\hat m}(t)$  not being able to perform task processing. In this case, the task will be offloaded to the cloud for processing. Let $R$ be the data rate of the backbone network, we can get the expected task processing delay:
\begin{equation}
	D^{pro}_{T_u(t)}=x_{\hat k,\hat m}(t)D^{edge}_{T_u(t)}+\big(1-x_{\hat k,\hat m}(t)\big)D^{BKB}_{T_u(t)},
\end{equation}
among them,
\begin{subequations}
	\begin{align}
		&D^{edge}_{T_u(t)}=\frac{w_{T_u(t)}}{f_{\hat k}},\\
		&D^{BKB}_{T_u(t)}=\frac{d_{T_u(t)}}{R},
	\end{align}
\end{subequations}
represent the computing delay at the edge and the transmission delay when offloading to the cloud, respectively. To reduce the bandwidth occupation of the backbone network, the tasks are expected to be processed at the edge as far as possible. Therefore, assuming that $R$ is a small value, offloading to the cloud will lead to a large transmission delay.

Added with the uplink delay above, the offloading delay can be obtained as
\begin{equation}
	D^{total}_{T_u(t)}=D^{UCN}_{T_u(t)}+D^{pro}_{T_u(t)}.
\end{equation}

\subsection{Problem Formulation}

The calculation of beamforming vector requires the exchange of channel information within the BS cluster through backhaul links. To reduce the bandwidth occupation of backhaul links, the size of BS cluster should be limited. The constraint of BS clustering size is
\begin{equation}
	1\leq\sum_{m\in \mathcal M}c_{u,m}(t)\leq B,\label{c1}
\end{equation}
where $B$ is the maximum size of BS cluster.We assume the total long term total caching cost can not exceed a threshold $Cost^{th}$, which can be expressed as
\begin{equation}
	\frac1T\sum_{t\in T}\sum_{m\in \mathcal{M}}c^{*}_{u,m}(t)Cost_m(t)<Cost^{th}.\label{c2}
\end{equation}
Moreover, for each BS, the caching and computing resources consumed by caching the services cannot exceed the resource capacity: 
\begin{subequations}
	\begin{align}
		\label{c4}
		&\quad \sum_{k\in \mathcal{K}}x_{k,m}(t)s_k\leq c_{u,m}(t)S_m,\forall m\in \mathcal M, \\
		\label{c3}
		&\quad \sum_{k\in \mathcal{K}}x_{k,m}(t)f_k\leq c_{u,m}(t)C_m,\forall m\in \mathcal M.
	\end{align}
\end{subequations}

We first consider the single-user offloading scenario where only one user generates the offloading tasks. The problem can be easily expanded to the multi-user scenario, which will be described in Section IV. Mathematically, the long-term delay minimization problem of user $u$ can be formulated as
\begin{subequations}\label{P}
	\begin{align}
		\mathrm{P}:
		&\quad \min_{\boldsymbol C(t),\boldsymbol X(t)}\frac1T\sum_{t\in T}D_{T_u(t)}^{total}\\
		s.t.\;
		\label{costC2}
		&\quad (9) -(11),\\
		\label{variC2}
		&\quad c_{u,m}(t)\in \{0,1\},\;x_{k,m}(t)\in [0,1],\forall k\in \mathcal K,\forall m\in \mathcal M.
	\end{align}
\end{subequations}

Where ${\boldsymbol X}(t)=[\boldsymbol x_m(t)|m\in\Phi_u(t)] $, ${\boldsymbol C}(t)=[\boldsymbol c_m(t)|\\  m\in\Phi_u(t)]$ are the service caching matrix and BS clustering matrix of user $u$, respectively. 

The first challenge that impedes the derivation of the optimal solution to the above problem is that the service caching cost constraint (10) makes the variables coupled in multiple time slots. Moreover, the existence of binary variable $c$ and continuous variable $x$ makes the problem become a mixed-integer nonlinear programming (MINLP) problem, which is NP-hard.

\section{Joint BS Clustering and Service Caching Optimization}
To overcome the above difficulties, we design an online algorithm called JO-CDSD. In particular, we first transform the long-term optimization problem into a time-decoupled instantaneous problem based on Lyapunov optimization and then decompose the instantaneous problem into two sub-problems. Then these sub-problems are solved alternately to obtain the near-optimal BS clustering and service caching strategy.

\subsection{Lyapunov-based Problem Transformation}
Based on the Lyapunov optimization\cite{Neely2010}, we construct a virtual caching cost queue $C(t)$ which represents the backlog of caching cost of the current slot\cite{xu2018joint,Hu2019,Duan2020}. We assume the initial state is $C(0)=0$, and the state transition of the queue can be written as
\begin{equation}\label{dynamic}
	C(t+1)=[C(t)+a(t)-b(t)]^{+},
\end{equation}
where $a(t)=\sum_{m\in \mathcal{M}}c^{*}_{u,m}(t)\big(\sum_{k\in \mathcal{K}}\xi_k s_k x_{k,m}(t)\big)$ and $b(t)=Cost^{th}$ represent the task arrival rate and service rate of the queue, respectively, and $[\cdot]^{+}$ represents $ \max\{\cdot,\;0\}$. The Lyapunov function can be obtained as $L(t)=1/2C(t)^2$, and the Lyapunov drift is $\Delta(t)=L(t+1)-L(t)$. According to the definition above, the Lyapunov drift can be rewritten as 
\begin{equation}\label{upper1}
	\begin{aligned}
		\Delta(t)&=\frac12C(t+1)^2-\frac12C(t)^2\\
		&\leq\frac12\big(C(t)+a(t)-b(t)\big)^2-\frac12C(t)^2\\
		&=Q(t)+C(t)\big(a(t)-b(t)\big),
	\end{aligned}
\end{equation}
where $Q(t)=1/2\big(a(t)-b(t)\big)^2$. The upper bound can be derived as:
\begin{equation}\label{upper2}
	Q(t)\leq \overline {Q(t)}=\frac12\big(\big(\sum_{m\in \mathcal{M}}c^{*}_{u,m}(t)\sum_{k\in\mathcal{K}}\xi_ks_k\big)^2+ (Cost^{th})^2\big).
\end{equation}
Based on the Lyapunov optimization framework, the long-term caching cost constraint (\ref{c2}) can be transformed into a drift minimization problem at each slot. Therefore, we can transform the long-term optimization problem to the instantaneous optimization problem with the optimization goal of \emph{ drift-plus-penalty}:
\begin{subequations}\label{P-lya}
	\begin{align}
		\mathrm{P-Lyapunov}:
		& \min_{\boldsymbol C(t),\boldsymbol X(t)}C(t)(a(t)-b(t))+VD_{T_u(t)}^{total}\\
		s.t.
		&\quad (\ref{c1}),(\ref{c4}),(\ref{c3}),(\ref{variC2}),
	\end{align}
\end{subequations}
where $V$ is a non-negative weight factor, which is selected according to the trade-off between the caching cost queue drift and the offloading delay.

As far as the problem $\mathrm{P-Lyapunov}$ is concerned, the numbers of variables and constraints of problem $\mathrm{P-Lyapunov}$ reach $M(K+U)$ and $1+M(2+U+K)$, respectively. If we adopt traditional searching algorithm like branch-and-bound (BnB)\cite{Hou2010} to solve the problem, the worst-case complexity will remain exponential, which makes it hard to obtain the solution in reasonable time especially when the numbers of BSs, users and types of services grow. Thus, in the next section, we will decompose the problem based on generalized Benders decomposition (GBD) to solve the problem $\mathrm{P-Lyapunov}$ with low complexity.

\subsection{GBD-based Problem Decomposition}

J. F benders\cite{2005Partitioning} proposed a mathematical solution framework for the complex variables programming problem ,which is called Benders decomposition. In Benders decomposition, when the complex variables are fixed, the remaining optimization problems will degenerate into linear problems which are easier to solve. Benders decomposition uses the cutting plane method to find the optimal value, and represents the solution of the degenerate problem as a function of the complex variables. A. M. Geoffrion\cite{geoffrion1972generalized} extended the Benders framework to suitable for general problems, which is called generalized Benders decomposition.

Based on the GBD framework, we propose the joint optimization algorithm JO-CDSD to solve $\mathrm{P-Lyapunov}$. Consider binary variable $c$ as the complex variable, we need to ensure the problem meets the requirement of GBD, that is, we can get a convex optimization problem of variable $x$ when fixing variable $c$. After analyzing $\mathrm{P-Lyapunov}$, the optimization goal consists of two parts: the Lyapunov drift and the Lyapunov penalty, which is shown as (17a) and (17b), respectively

\begin{subequations}
\begin{align}
	&C(t) (\Xi^T \boldsymbol X(t) \boldsymbol C(t)^T - Cost^{th})  \\
	&\boldsymbol V (D^{UCN}_{T_u(t)} + \max_{m \in \boldsymbol M}\{\boldsymbol C(t) \odot  o^T\boldsymbol X(t) \}   (D_{T_u(t)}^{UCN}+D_{T_u(t)}^{edge})  )
\end{align}
\end{subequations}

It can be seen that these two parts are in a competitive relationship for the variable $x$: when $x$ increase, the Lyapunov drift is non-decreasing and the the Lyapunov penalty in the current slot is non-increasing. Since the Lyapunov penalty is a concave function of $x$, the optimization goal obtained by the direct addition of these two parts is non-convex with respect to $x$,  thus it does not satisfy the decomposition requirement of GBD. Therefore, we add a service caching probability constraint to make sure that at least one BS in the BS cluster will cache the requested service. As a result, the problem can be converted into a convex  problem of variable $x$ after fixing $c$. Adding this constraint has no effect on the delay optimality, but it may cause the problem to be infeasible. In this case, we can relax the service caching probability to $[0,1]$. The service caching probability will decrease from $1$ until the problem is feasible. By adding the service caching probability constraint of the BS cluster,  the instantaneous optimization problem (16) can be rewritten as 
\begin{subequations}\label{P lya relax}
	\begin{align}
		&\min_{\boldsymbol C(t),\boldsymbol X(t)}C(t)\big(a(t)-b(t)\big)+V\big(D^{UCN}_{T_u(t)}+D^{edge}_{T_u(t)}\big)\\
		s.t.
		\label{maxC}
		&\quad \max_{m\in\mathcal M}\{\boldsymbol C(t)\odot(\boldsymbol o^T\boldsymbol X(t))\}=1,\\
		\label{vec-consS}
		&\quad \boldsymbol s^T\boldsymbol X(t)\leq\boldsymbol C(t)\boldsymbol S,\\
		\label{vec-consC}
		&\quad \boldsymbol f^T\boldsymbol X(t)\leq\boldsymbol C(t)\boldsymbol C,\\
		&\quad \boldsymbol C(t)\in \mathcal F_c,\;\boldsymbol X(t)\in \mathcal F_x.
	\end{align}
\end{subequations}

where $\boldsymbol o$ is the request service type vector. When $c$ is fixed, the goal is to optimize the Lyapunov drift, which is a convex function of $x$.  The constraint (18b) is a function in the form of $h(g(x))$, where $h(x) =\max_i{x_i}$, $g(x)=Ax$ are the convex function of $x$. According to the convexity preserving property of composite functions, $h(g(x))$ is convex function when $g(x)$, $h(x)$ are both convex function and $h(x)$ is non-decreasing. Therefore, (18b) is also a convex function of $x$. To sum up, optimization problem (18) satisfies the assumption of GBD. Next, we decompose the problem (18) into service caching problem (P-primal) and BS clustering problem (P-master), which is shown in lemma 4.1.

\textbf{Lemma  4.1} By fixing the BS clustering strategy $\boldsymbol C(t) = \boldsymbol{\overline{C}}(t)$, the optimization problem (18) can be decomposed into two sub-problems
\begin{subequations}\label{primal_lemma}
	\begin{align}
		\mathrm{P-primal}:
		&\quad \begin{array}{l}\min_{\boldsymbol X(t)\in{\mathcal F}_x}C(t)(\boldsymbol\Xi^T\boldsymbol X(t)\overline{\boldsymbol C}(t)^T-\\Cost^{th})+V(D_{T_u(t)}^{UCN}+D_{T_u(t)}^{edge})\end{array}\\
		s.t.
		\label{cons_lemma1}
		&\quad \max_{m\in\mathcal M}\{\overline{\boldsymbol C}(t)^T\odot (\boldsymbol o^T\boldsymbol X(t))\}=1,\\
		\label{cons_lemma2}
		&\quad \boldsymbol s^T\boldsymbol X(t)\leq\overline{\boldsymbol C}(t)^T\boldsymbol S,\\
		\label{cons_lemma3}
		&\quad \boldsymbol f^T\boldsymbol X(t)\leq\overline{\boldsymbol C}(t)^T\boldsymbol C.
	\end{align}
\end{subequations}

and

\begin{subequations}\label{p-master}
	\begin{align}
		\mathrm{P-master}:
		&\quad \min_{\boldsymbol C(t)}v\big(\boldsymbol C(t)\big)\\
		s.t.
		&\quad \boldsymbol C(t)\in \mathcal F_c\cap Y.
	\end{align}
\end{subequations}

the optimal value function $v$ and feasibility set $\boldsymbol Y$ of complex variables $c$ are defined as follows: 

\begin{equation}
	v(\boldsymbol C(t))\equiv\inf_{ X(t)}\{C(t)(a(t)-b(t))\},\;s.t.(\ref{cons_lemma1})-(\ref{cons_lemma3}),
\end{equation}

\begin{equation}
\begin{array}{l}Y\equiv C(t):\exists X(t)\in{\mathcal F}_x,
	\\
	\phantom{=\;\;}
	\begin{array}{lr}(\boldsymbol s^TX(t))^T\leq\boldsymbol C(t)^T\odot\boldsymbol S\\(\boldsymbol f^TX(t))^T\leq\boldsymbol C(t)^T\odot\boldsymbol C\\\max_{m\in\mathcal M}\{C(t)\odot(\boldsymbol o^T\boldsymbol X(t))\}=1\end{array},\end{array}
\end{equation}

\textbf{Proof}: (20) is a convex optimization problem of $x$, and the definition of the feasibility set also meets the assumptions in \cite{geoffrion1972generalized}. Therefore, the equivalence proof between the instantaneous optimization problem (18) and the above decomposition process can be given by theorem 2.1 in \cite{geoffrion1972generalized}.

The Lagrange functions of $\mathrm{P-primal}$ can be derived as
\begin{equation}
	\begin{split}
	\begin{array}{l}\begin{array}{l}\mathcal L(\boldsymbol X(t);\overline{\boldsymbol C}(t),\boldsymbol\mu)=\boldsymbol C(t)(a(t)-b(t))+VD_{T_u(t)}^{total}\\+\boldsymbol\mu_1^T((\boldsymbol s^T\boldsymbol X(t){)^T-\overline{\boldsymbol C}(t)^T\odot\boldsymbol S)+\boldsymbol\mu_2^T((\boldsymbol f^T\boldsymbol X(t))^T}\end{array}\\-\overline{\boldsymbol C}(t)^T\odot\boldsymbol C)+\mu_3(\max_{m\in\mathcal M}\{\overline{\boldsymbol C}(t)\odot(\boldsymbol o^T\boldsymbol X(t))\}-1)\end{array},
	\end{split}
\end{equation}

\begin{equation}
	\begin{split}
		\begin{array}{l}\begin{array}{l}\widetilde{\mathcal L}(\boldsymbol X(t);\overline{\boldsymbol C}(t),\boldsymbol\lambda)=\boldsymbol\lambda_1^T((\boldsymbol s^T\boldsymbol X(t))^T-\overline{\boldsymbol C}(t)^T\odot\boldsymbol S)\\+\boldsymbol\lambda_2^T((\boldsymbol f^T\boldsymbol X(t))^T-\overline{\boldsymbol C}(t)^T\odot\boldsymbol C)\end{array}\\+\lambda_3(\max_{m\in\mathcal M}\{\overline{\boldsymbol C}(t)\odot(\boldsymbol o^T\boldsymbol X(t))\}-1),\end{array}
	\end{split}
\end{equation}
where the multiplier item corresponds to the constraint in the question (\ref{primal_lemma}), and there are $\mu_i,\lambda_i\geq 0$ and $\boldsymbol \mu, \boldsymbol \lambda\in \mathbb R^{2K+ 1}$.  Based on the Lagrange functions, we can rewritten the P-master problem as
\begin{subequations}
	\begin{align}
		P-master:
		&\quad \min_{ C(t),d_0}d_0\\
		s.t.
		&\quad d_0\geq \inf_{ X(t)\in\mathcal F_x}\mathcal L( X(t); C(t),\boldsymbol \mu), \forall \boldsymbol \mu \geq 0,\\
		&\quad 0\geq \inf_{X(t)\in\mathcal F_x}\tilde{\mathcal L}( X(t); C(t),\boldsymbol \lambda),\forall \boldsymbol \lambda\in\Lambda.
	\end{align}
\end{subequations}
where
\begin{equation}
	\Lambda \equiv\left\{\boldsymbol \lambda\in \mathbb R^{2M+1}:\boldsymbol \lambda\geq 0\;and\;\sum_{i=1}^{2M+1} \lambda_i=1\right\}.
\end{equation}

\subsubsection{The Solution of P-Primal Problem}
Since $\mathrm{P-primal}$ is a convex problem, the constraints in $\mathrm{P-master}$ can be written as:

\begin{equation}
	\begin{array}{l}d_0\geq\mathcal L^\ast(\boldsymbol C(t),\boldsymbol X^{ }(t)^{(\tau_1)},\boldsymbol\mu^{(\tau_1)})\;,\\\tau_1\in\{\tau\vert if\;(\ref{primal_lemma})^{(\tau)}\;is\;feasible.\},\forall\boldsymbol\mu\geq0,\end{array}
\end{equation}
\begin{equation}
	\begin{array}{l}0\geq\widetilde{\mathcal L}^\ast(\boldsymbol C(t),\boldsymbol X^{ }(t)^{(\tau_2)},\boldsymbol\lambda^{(\tau_2)})\;,\\\tau_2\in\{\tau\vert if\;(\ref{primal_lemma})^{(\tau)}\;is\;infeasible.\},\forall\boldsymbol\lambda\in\Lambda,\end{array}
\end{equation}

where $\{ \boldsymbol X(t)^{(\tau)},\boldsymbol \mu^{(\tau)},\boldsymbol \lambda^{(\tau)}\}$ denotes the optimal solution and the optimal multiplier vector obtained by solving problem  (\ref{primal_lemma}) in the $\tau$-th iteration.

$\mathrm{P-primal}$ can be solved by using the convex optimization toolkit. During each iteration, the solution of $\mathrm{P-primal}$ will be added to $\mathrm{P-master}$ problem as the new constraints. From the definition of the set $Y$, it can be seen that when fixing $c$, $\mathrm{P-primal}$ is not always feasible.

If $\mathrm{P-primal}$ is feasible, the optimal solution and the bounded optimal value can be obtained to generate the constraints 
\begin{equation}
	d_0\geq \mathcal L^{*}( \boldsymbol C(t), \boldsymbol X(t)^{(\tau)},\boldsymbol \mu^{(\tau)}).
\end{equation}
Otherwise, the constraints in $\mathrm{P-primal}$ problem cannot be satisfied. However, we can obtain a near-optimal solution with the minimum damage to the constraints by solving the following problem: 

\begin{subequations}\label{infeasible}
	\begin{align}
		\mathrm{P-}&\mathrm{primal(infeasible)}:\quad \min_{ \boldsymbol X(t)\in\mathcal F_x,\alpha}\alpha \\
		s.t.
		&\quad \alpha>0,\\
		&\quad \sum_{k\in\mathcal K}x_{k,m}(t)s_k-\overline c_{u,m}(t)S_m\leq\alpha,\forall m\in\mathcal M,\\
		&\quad \sum_{k\in\mathcal K}x_{k,m}(t)f_k-\overline c_{u,m}(t)C_m\leq\alpha,\forall m\in\mathcal M.
	\end{align}
\end{subequations}
The optimal solution $(\boldsymbol X(t)^{(\tau)},\boldsymbol \lambda^{(\tau)})$ of the above problem can be used to generate the infeasible constraint:
 \begin{equation}
	0\geq \tilde{\mathcal L}^{*}(\boldsymbol C(t),\boldsymbol X(t)^{(\tau)},\boldsymbol \lambda^{(\tau)}),
\end{equation}

\subsubsection{The Solution of P-Master Problem}
$\mathrm{P-master}$ is a 0/1 programming problem, and the size of the decision space is $\sum_{b=1} ^B\binom{M}{b}$, hense cannot be solved by the traditional dynamic programming algorithm within affordable complexity. To overcome the above difficulties, we propose the BS clustering algorithm based on the Gibbs Sampling to find the near-optimal solution of $\mathrm{P-master}$.

We define the Lagrange function of $\mathrm{P-master}$ as 
\begin{equation}
\begin{array}{l}F(C(t),d_0,\boldsymbol\beta)=d_0+\sum_{\tau_1\in\{\tau\vert if\;(19)^{(\tau)}\;is\;feasible.\}} 

\\   \beta^{(\tau_1)}(\mathcal L^\ast(\boldsymbol C(t),\boldsymbol X(t)^{(\tau)},\boldsymbol\mu^{(\tau)})-d_0).\end{array}
\end{equation}

Consider such a graph: all idle BSs are regarded as vertices, and the clustering state of BS $m$ is $c_{u,m}\in\{0,1\}$. Taking the Lagrange function of $\mathrm{P-master}$ as the optimization objective, the probability distribution of the state transition of vertex $m$ can be obtained as 

\begin{footnotesize}
\begin{equation}
	\begin{aligned}
		&\pi_m(\boldsymbol c_{u,\overline m})=\pi_m(c_{u,m}|\boldsymbol c_{u,\overline m})\\
		&=\frac{exp\left(-F(c_{u,m},\boldsymbol c_{u,\overline m})/\varphi\right)}{exp\left(-F(c_{u,m}=0,\boldsymbol c_{u,\overline m})/\varphi\right)+exp\left(-F(c_{u,m}=1,\boldsymbol c_{u,\overline m})/\varphi\right)},
	\end{aligned}
\end{equation}
\end{footnotesize}
where $\boldsymbol c_{u,\overline m}$ denotes the state of other vertices except vertex $m$, i.e, the clustering state matrix of other BSs. $\varphi>0$ is the temperature parameter of the Gibbs sampling. When $\varphi\rightarrow 0$ and the sampling period tends to infinity, the system will converge to the optimal value. In the probability distribution, the calculation of the optimization target $F$ also depend on the value of $d_0$. When the clustering variable $c_{u,m}$ of the vertex is known, the problem of minimizing $F$ becomes an unconstrained convex optimization problem of minimizing $d_0$, which can be solved easily. Thus we can obtain the optimal $F$ of BS $m$ in the current clustering state. Moreover, we can sample according to the above probability distribution to complete the state transition of BS $m$. 

After updating the clustering status of BS $m$, the clustering status of the next BS is updated by randomly select the next vertex. The exploration probability of state updating is obtained by

\begin{equation}
	\begin{aligned}
		\eta&=\frac{exp(-\tilde F/\varphi)}{exp(-\tilde F/\varphi)+exp(-F/\varphi)}\\
		&=\frac{1}{1+exp(min\{(\tilde F-F),\rho\}/\varphi)},
	\end{aligned}
\end{equation}

where $\rho>0$ is a parameter to adjust the exploration probability. When the new target value is much larger than the current target value, the difference between the two target values can be limited to $\rho$ to ensure a large exploration probability. We use $F$ and $\tilde F$ to denote the current state and the new state, respectively. During the BS clustering state transition, the probability to accept the new state is $\eta$, while the probability to remain the current state is $1-\eta$. The BS clustering algorithm to solve $\mathrm{P-master}$ based on the Gibbs Sampling is shown in \textbf{Algorithm \ref{clustering}}.

\begin{algorithm}[tbp]
	\caption{BS Clustering algorithm based on Gibbs Sampling}
	\label{clustering}
	\begin{algorithmic}[1]
		\STATE {\textbf{Initialization}
		\STATE $\boldsymbol C=\boldsymbol 0,F=+\infty,iter=1$ \;
		Determine whether each BS is idle, and use idle BSs as vertices to form a graph $\mathbb G$ \;}
		\FOR{$m$ in $V_{\mathbb G}$}
		\STATE Solve the optimization problem in the current state $\min_{d_0}F(d_0,\boldsymbol \beta|\boldsymbol C(t))$ and obtain the optimal solution $\tilde F$;\;
		\STATE Obtain the new state $\tilde c_{u,m}$ by randomly select the next vertex. 
		\STATE Solve the optimization problem in the new state $\min_{d_0}F(d_0,\boldsymbol \beta|\tilde {\boldsymbol C}(t)) $ and obtain the optimal solution $\tilde F$;\;
		\STATE Calculate $\eta$, update $c_{u,m}=\tilde c_{u,m}$ with probability $\eta$\
		\STATE $iter+=1$
		\STATE check whether the iteration stop condition is satisfied, such as when the iteration step reaches the maximum iteration step\; 
		\ENDFOR
		\STATE \textbf{output:}  BS clustering variable in current status $\mathbf C$ 
\end{algorithmic}
\end{algorithm}

In summary, the JO-CDSD algorithm based on the GBD is shown in \textbf{Algorithm \ref{my gbd}}.

\subsection{Extend to the Multi-User Scenario}
we have discussed the joint BS clustering and service caching optimization when a specific user generates an offloading request. When multiple users generate offloading requests at the same time, the optimization of BS clustering and service caching strategy will become more complex. In this subsection, we will extend our JO-CDSD algorithm to the multi-user scenario.

In the multi-user scenario, the division of BS clusters can be divided into the following cases: 
\begin{itemize}
    \item \textbf{(a)}. All users share a same BS cluster. 
    \item \textbf{(b)}. The BS clusters of users are completely different.
    \item  \textbf{(c)}. The BS clusters of different users are partially coincide 
\end{itemize}

In the case \textbf{(b)}, the problem can be decomposed into multiple single-user optimization problems, and the optimal BS clustering and service caching strategy can be obtained by using \textbf{Algorithm 2}. In the case \textbf{(c)}, the BSs in the overlapping area can be regarded as a BS cluster serving multiple users, i.e., case \textbf{(a)}, and the BSs in the non-overlapping area can be regarded as a BS cluster serving a single user. In this paper, we mainly discuss the most common case \textbf{(a)}, i.e., multiple users sharing the same BS cluster and generating the offloading requests at the same time. We assume that the offloading requests for the same type of services can be regarded as a request from one user and the corresponding tasks can be merged into one task too.

\begin{algorithm}[tbp]
	\caption{JO-CDSD Algorithm}
	\label{my gbd}
	\begin{algorithmic}[1]
	\STATE {\textbf{Initialization:} $\epsilon>0$, select the initial value of the complex variable $\overline {\boldsymbol C}\in\mathcal F_c$ and upper bound $UBD=+\infty$ lower bound $LBD=-\infty$ \; }
		\WHILE{$\tau<\tau^{max}$}
		\STATE 	Solve the P-primal problem (\ref{primal_lemma})
			\IF{ problem (\ref{primal_lemma}) is feasible}
	\STATE Check whether the optimal value is bounded. If it is bounded, check whether the optimal value meets the iteration stop condition: $minimized(\ref{primal_lemma})-LBD^{(\tau)}\leq\epsilon$. Otherwise get the optimal multiplier vector $\overline{\boldsymbol \mu}$ and the optimal solution $\overline { X}(t)$, and get the constraint about $ C (t)$:  $d_0\geq \mathcal L^{*}(\boldsymbol C(t),\boldsymbol{\overline{X}}(t)^{(\tau)},\overline{\boldsymbol \mu}^{ (\tau)})$. Let $\boldsymbol \mu^{(\tau)}=\overline{\boldsymbol \mu}$, $UBD=\min\{UBD,minimized(\ref{primal_lemma})\}$\; 
	\ELSE
	\STATE Problem (\ref{primal_lemma}) is not feasible. Solve the problem (\ref{infeasible}) to obtain the service caching variable $\boldsymbol{\overline {X}}(t)$ which  minimizes the destruction of the constraints. Obtain the optimal multiplier vector $\overline {\boldsymbol \lambda}_1,\overline {\boldsymbol{\lambda}}_2\in\Lambda$ and constraint $0\geq \tilde{\mathcal L}^{*}(\boldsymbol C(t),\overline{  X}(t),\overline{\boldsymbol \lambda})$, let $\boldsymbol \lambda^{(\tau)}=\overline {\boldsymbol \lambda}$\; 
	\ENDIF
		\STATE  Solve $\mathrm{P-master}$ according to \textbf{Algorithm \ref{clustering}} and get the optimal solution $\boldsymbol {\hat{C}}(t)$ or sub-optimal solution $\hat d_0$. Let current lower bound $LBD^{(\tau)}=\hat d_0$, and check whether the iteration stop condition is satisfied, i.e., $UBD^{(\tau)}-LBD^{(\tau)}\leq\epsilon$.\; 
		\STATE Let $\boldsymbol{\overline {C}}(t)=\boldsymbol{\hat{C}}(t),\tau +=1$\;
		\ENDWHILE	
	 \STATE \textbf{output:} The optimal BS clustering variable and service caching variables at the current slot $\boldsymbol X(t)$ and $\boldsymbol C(t)$
	\end{algorithmic}
\end{algorithm}

We assume the set of users that sharing a BS cluster are $\mathrm B_0$, and the offloading requests are described as $T_u(t) : [d_{T_u(t)}, w_{T_u(t)}, \boldsymbol o_{T_u(t)}] $, which denote the data size, computing workload, and service type vector, respectively. We assume that each user in $\mathrm B_0$ request only one service in each time slot, and the type of services requested by multiple users at each slot can be different. The BS clustering variable and the service caching variable of users in $\mathrm B_0$ can be denoted as $\boldsymbol C^{\mathrm B_0}(t)$, and $\boldsymbol X^{\mathrm B_0}(t)$, respectively. Similar to (16), the instantaneous BS clustering and service caching problem in multi-user scenario can be obtained based on the Lyapunov optimization

\begin{subequations}\label{P2 Lyapunov}
	\begin{align}
	\mathrm{P2-}&\mathrm{Lyapunov}:
		\min _{\boldsymbol C^{B_{0}}(t), \boldsymbol X^{B_{0}}(t)}\frac{1}{\left|\boldsymbol{B}_{0}\right|} \sum_{u \in B_{0}} \nonumber\\
		&\left(\begin{array}{l}
\boldsymbol C(t)\left(\Xi^{T} \boldsymbol{X}^{B_{0}}(t) \boldsymbol C^{B_{0}}(t)^{T}-\operatorname{Cost}^{t h}\right)+ \\
V\left(\frac{d_{T_{u}(t)}}{r_{B_{0}}(t)}+\max _{m \in \mathcal{M}}\left\{\boldsymbol C^{B_{0}}(t) \odot\left(o_{T_{u}(t)}^{T} X^{B_{0}}(t)\right)\right\}\right. \\
\left.\left(D_{T_{u}(t)}^{e d g e}-D_{T_{u}(t)}^{B K B}\right)+D_{T_{u}(t)}^{B K B}\right)
\end{array}\right)\\
		s.t.
		\label{maxC}
		&\quad  \boldsymbol s^T \boldsymbol X^{\mathrm B_0}(t) \leq \boldsymbol C^{\mathrm B_0}(t) \boldsymbol S,\\
		\label{vec-consS}
		&\quad  \boldsymbol{f}^T \boldsymbol X^{\mathrm B_0}(t) \leq \boldsymbol C^{\mathrm B_0}(t) \boldsymbol{C}  ,\\
		\label{vec-consC}
		&\quad \boldsymbol C^{\mathrm B_0}(t) \in \mathcal F_c,\\
		&\quad X^{\mathrm B_0}(t) \in \mathcal F_x.
	\end{align}
\end{subequations}

After fixing the complex variable $\boldsymbol C^{\mathrm B_0}(t)$, the problem is still non-convex to $\boldsymbol X^{\mathrm B_0}(t)$. Similar to the previous section, we transform it into a convex problem by adding a constraint

\begin{subequations}\label{primal}
	\begin{align}
		\mathrm{P2-}&\mathrm{Lyapunov(\Theta)}:\\
		&\quad \min _{\boldsymbol C^{R_{0}}(t), \boldsymbol X^{B_{0}(t)}}\ C(t)\left(\Xi^{T} \boldsymbol{X}^{B_{0}}(t) \boldsymbol C^{B_{0}}(t)^{T}-\operatorname{Cost}^{t h}\right) \nonumber
		\\
		&+\frac{V}{\left|B_{0}\right|} \sum_{u \in B_{0}} \frac{d_{T_{u}(t)}}{r_{B_{0}}(t)}+V \Theta\\
		s.t.
		\label{cons1}
		&\quad \frac{1}{\left|B_{0}\right|} \sum_{u \in B_{0}}\left(\max _{m \in \mathcal{M}}\left\{\boldsymbol C^{B_{0}}(t) \odot\left(o_{T_{u}(t)}^{T} X^{B_{0}}(t)\right)\right\} \right.\\ \nonumber &\left.\left(D_{T_{u}(t)}^{BKB}-D_{T_{u}(t)}^{edge}\right)-D_{T_{u}(t)}^{BKB}\right)=-\Theta ,\\
		&\quad (31c) - (31f).
	\end{align}
\end{subequations}

where $\Theta$ is a hyper-parameter to balance the optimality and feasibility of the problem. When the computing resources are sufficient, all the services requested by users can be cached on the BS cluster. Therefore, we can obtain the minimum average offloading delay
\begin{equation}
	\underline{\Theta} = \frac{1}{|\mathrm B_0|} \sum_{u \in \mathrm B_0} {D^{edge}_{T_u(t)}}
\end{equation}

If the BS cluster does not have enough resources, all the tasks will be offloaded to the cloud for processing. In this case, the maximum average offloading delay can be written as
\begin{equation}
	\overline \Theta = \frac{1}{|\mathrm B_0|} \sum_{u \in \mathrm B_0} {D^{BKB}_{T_u(t)}}
\end{equation}

To make a trade-off between optimality and feasibility, the value of $\Theta$ needs to be determined carefully. As a result, we propose a dichotomy-based GBD joint otimization algorithm to get the optimal $\Theta$ as well as the  corresponding BS clustering and service caching strategies, which is shown in \textbf{Algorithm 3}.

\begin{algorithm}[tbp]
	\small
	\caption{Dichotomy-based JO-CDSD Algorithm}
	\label{ gbd2}
	\begin{algorithmic}[1]
	\STATE  \textbf{Initialization:} $iter=1$ ,$iter^{max}>1$,$\epsilon>0$.Set the upper and lower bounds of $\Theta$ as $\overline \Theta$, $\underline{\Theta}$, $\Theta_{mid} = \frac{1}{2} (\overline{\Theta} + \underline{\Theta}) $, $\Theta_{mid}^{old} = \Theta_{mid}$
		\WHILE{$iter<iter^{max}$}
		\STATE 	Solve the problem (36) by JO-CDSD algorithm(\textbf{Algorithm 2})
			\IF{the problem (36) is feasible}
	\STATE  Calculate $\Theta_{mid}^{new} = \frac{1}{2} (\overline{\Theta} + \underline{\Theta})$ and check whether the iteration stop condition is satisfied: i.e $|\Theta_{mid}^{new} - \Theta_{mid}^{old} | < \epsilon$. Otherwise, update the upper bound $\overline{\Theta} = \Theta_{mid}^{old} $\; 
	\ELSE
	\STATE The problem is not feasible. Update the lower bound $\underline{\Theta} = \Theta_{mid}^{old} $. Calculate $\Theta_{mid}^{new} =  \frac{1}{2} (\overline{\Theta} + \underline{\Theta})$\; 
	\ENDIF
		\STATE $iter+=1$\;
		\ENDWHILE	
	 \STATE \textbf{output:} The optimal BS cluster division variable $\boldsymbol X^{\mathrm B_0}(t)$ and service caching variables $\boldsymbol C^{\mathrm B_0}(t)$
	\end{algorithmic}
\end{algorithm}

\subsection{Computational Complexity Analysis}
In \textbf{Algorithm 1}, Assuming that the maximum iteration steps is $\kappa$, thus the complexity of the algorithm is $O(\kappa UM)$. For problem (19), there are totally $KM$ optimization variables and $3M$ convex constraints. Thus the complexity of problem (19) would be $O(3KM^2)$.  For problem (30),  there are $KM$ optimization variables and $2M+1$ convex constraints, thus the complexity of problem (19) is given by $O(2KM^2+KM)$. Thus the worse-case computational complexity of \textbf{Algorithm 2} is given as $O(\tau^{max}(\kappa U+K)M + 5KM^2))$. Similarly, the complexity of \textbf{Algorithm 3} is calculated as $O(iter^{max}\tau^{max}((\kappa U+K)M +5KM^2))$.

\section{PERFORMANCE EVALUATION}
In this section, extensive simulations are provided to evaluate the effectiveness of the proposed JO-CDSD algorithm. We conduct our simulation on MATLAB R2021b, with 2.9 GHz Intel Core CPU i7 and 32 GB RAM. The simulation code is available on \href{https://github.com/qlt315/JO-CDSD}{https://github.com/qlt315/JO-CDSD}.

\begin{table}[htbp]\label{t2}
	\centering
	\caption{parameter settings}
	\label{t2}
	\begin{tabular}{cc}
		\toprule  
		parameter  &value \\
		\midrule  
            Number of BS $M$ & 10 \\
            Number of antennas of each BS $A$ &3 \\
		Maximum number of BS cluster $B$  &3 \\
            System Bandwidth $W$ &10 MHz \\
            Data size of offloading task $d_{T_u(t)}$  &[10,10*k] MB \\
            Computing workload of task $w_{T_u(t)}$  &[0.1,0.1*K] GHz \\
		Computing resources of BS $m$ $S_m$   &3 GHz \\
		Caching resources of BS $m$ $C_m$   & 3 Gbit  \\
		Requirement caching resources for service  $s_k$  &3 Gbit \\
		Requirement computing resources for service $f_k$ &0.3 GHz\\
		Service caching cost factor (per GB of data) $\xi_k$ &$[0.1,0.1*K]$\\
		Distribution of the request service type  &$\mathrm{Zipf}(0.5,K)$\\
            Data rate of backbone network $R$  & 0.05 Gbps \\
		\emph{Drift-plus-penalty} Weights $V$ &5\\
		Maximum number of GBD iterations  &2000\\
		GBD iteration stop error  &1e-4\\
		Gibbs maximum number of iterations  &2000\\
		Gibbs temperature parameters$\varphi$ &$[0.8-0.01]$\\
		Gibbs explores probability parameters $\rho$ &0.5\\
		Dichotomy maximum number of iterations $iter$ &10 \\
		GBD iteration stop error $\epsilon$ &1e-2\\
		
		\bottomrule 
	\end{tabular}
\end{table}
 In the single-user scenario, we set the number of BS $M=10$ and the number of antennas of each BS $A=3$. All BSs are randomly distributed within the range of $[0.1,1]$ km centered on the target user. We set the system bandwidth $W=10$ MHz and the data rate of the backbone network $R=0.05$ Gbps. The computing resources and caching resources of BS $m$ are $C_m=3$ GHz and $S_m=3$ Gbit, respectively. The data size and the computing requirement of task $d_{T_u(t)}$  and $w_{T_u(t)}$ are distributed in $[10,10*k]$ Mbit and $[0.1,0.1*K]$ GHz, respectively.  We adopt the number of service types $K=6$ and the distribution of service type requests by users follows the Zipf distribution Zipf$(0.5, K)$. The path loss of the channel is $128.1+37.6\rm{log}(distance)$ dB. In the multi-user scenario, we set the number of users as $3$,  and the transmission signals of users are orthogonal to each other. The main parameters are summarized in TABLE I.

To verify the feasibility and effectiveness of the proposed JO-CDSD algorithm, we compare the JO-CDSD algorithm with three benchmarks:
\begin{itemize}

\item{\textbf{Instant optimal}: Through traversing every possible BS clustering vector to solve the optimization problem after fixing the BS clustering variable. During the traversal process, the optimal BS clustering and service caching solutions can be obtained by recording the optimal solution of each step. Although instant can get the optimal solution, at the cost of high complexity due to the traversal searching method. In a network with $M$ BSs, the computational complexity of instant optimal will become $O(2^M)$).}

\item{\textbf{Uplink optimal}\cite{chen2021user}: The optimization of BS clustering and service caching are independent of each other. The optimal BS clustering is obtained by minimizing the uplink delay and the optimal service caching is obtained by minimizing the average total delay. }

\item{\textbf{Block descent}: Based on the block-by-block optimization method\cite{huynh2021joint}, the large-scale optimization problem can be decomposed into multiple blocks. By fixing other blocks and optimizing the upper bound of one block at a time, the global solution can be obtained alternately.}

\end{itemize}
	\begin{figure*}[htbp]
    \centering
    \captionsetup{justification=centering}
	  \subfloat[Convergence performance of Gibbs\protect\\ sampling clustering algorithm]{
       \includegraphics[width=0.45\linewidth]{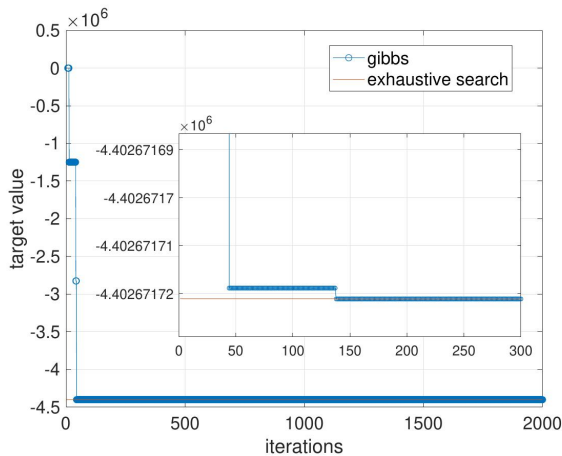}}
    \label{1a}
	  \subfloat[Convergence performance of JO-CDSD algorithm]{
        \includegraphics[width=0.45\linewidth]{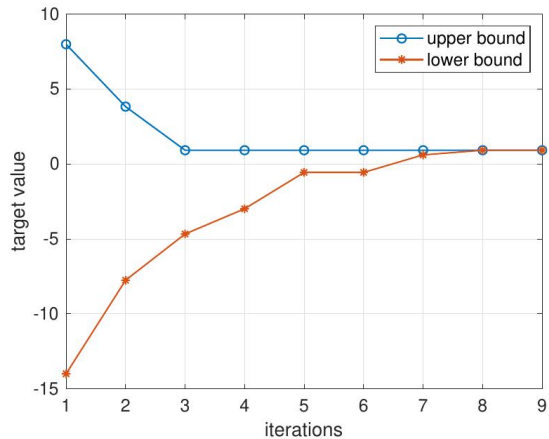}}
    \label{1b}
	  	  \caption{Convergence performance of (a) Gibbs sampling clustering algorithm and (b) JO-CDSD algorithm}
	  \label{fig1} 
	\end{figure*}

\subsection{Convergence Analysis}

We first evaluate the convergence performance of our proposed algorithm, including
Gibbs sampling clustering algorithm and JO-CDSD algorithm.

In Fig. 2(a), we evaluate the target value $\tilde F$ of the Gibbs sampling clustering algorithm (\textbf{Algorithm 1}) with an exhaustive search method. The exhaustive search method can obtain the optimal clustering decision by enumerating every possible BS clustering decision, but it has very high computational complexity. It can be seen that the target value of the Gibbs sampling clustering algorithm decreases rapidly in the first 50 iterations and converge to the optimal value in about 150 iterations. Compared with the exhaustive search method (1024 possible clustering combinations in a system with 10 BSs), the searching complexity of Gibbs sampling-based BS clustering is greatly reduced.

Fig. 2(b) shows the convergence performance of the JO-CDSD algorithm (\textbf{Algorithm 2}). It shows that the upper bound of the target value $d_0$ decreases rapidly and remains unchanged in the 4-th iteration, while the lower bound of the target value increase gradually. The difference between the upper and lower bound will decrease to the iteration stop error within 10 iterations. Although we use the iterative algorithm to solve the P-master problem, which will increase the complexity exponentially, the convergence performance results show that the GBD framework can converge quickly.

 \begin{figure*}[tbp]
    \centering
    \captionsetup{justification=centering}
	  \subfloat[Time average delay]{
       \includegraphics[width=0.3\linewidth]{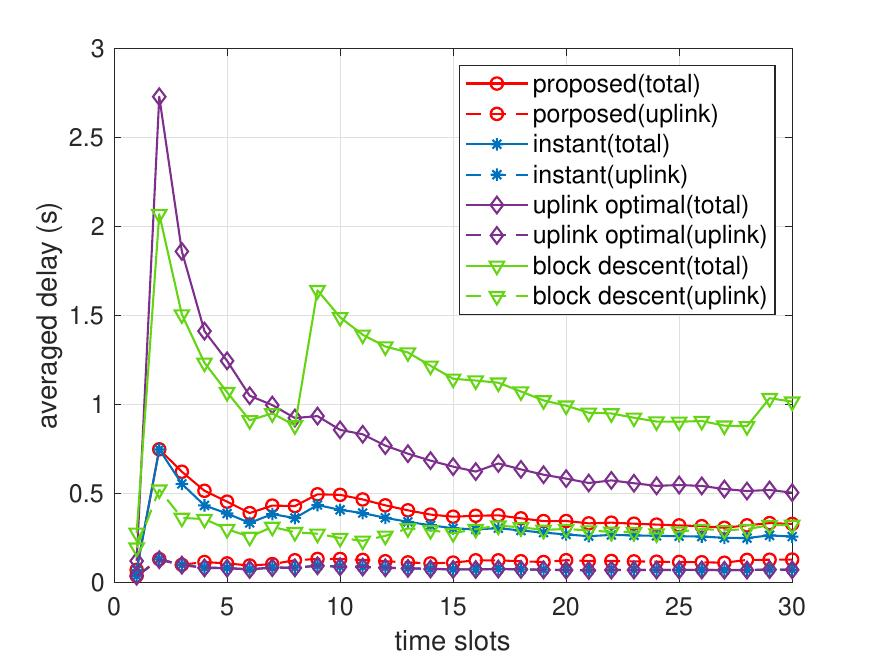}}
    \label{1a}
	  \subfloat[Time average cost]{
        \includegraphics[width=0.3\linewidth]{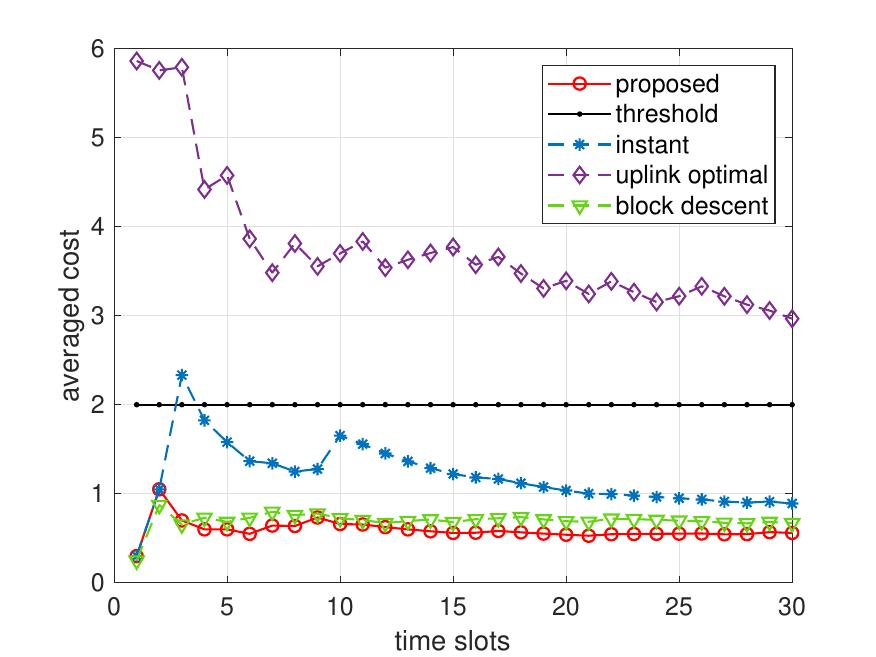}}
    \label{1b}
	  \subfloat[Time average delay with \protect\\different $cost^{th}$]{
        \includegraphics[width=0.3\linewidth]{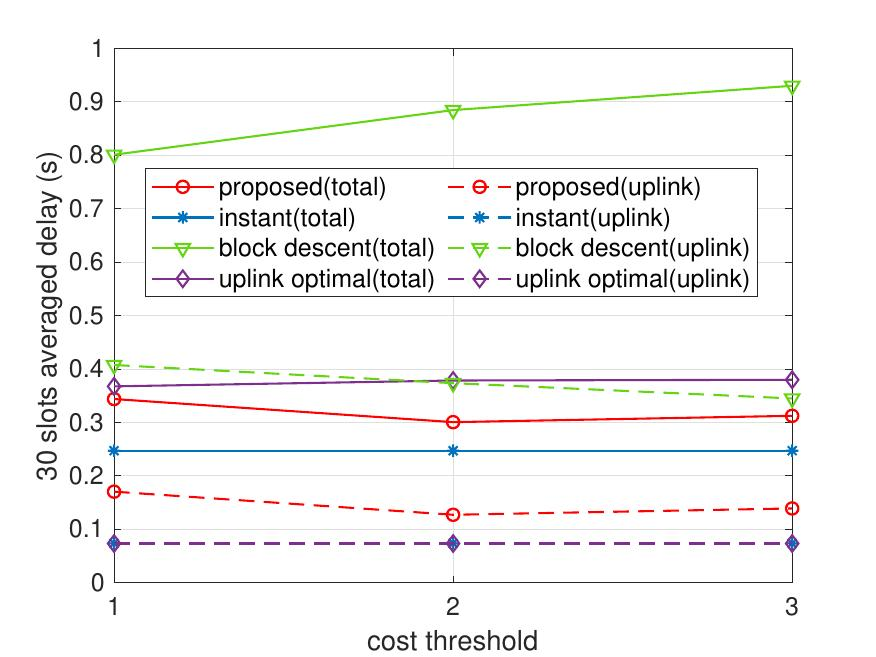}}
    \label{1c}
     \subfloat[Time average cost with \protect\\different $cost^{th}$]{
       \includegraphics[width=0.3\linewidth]{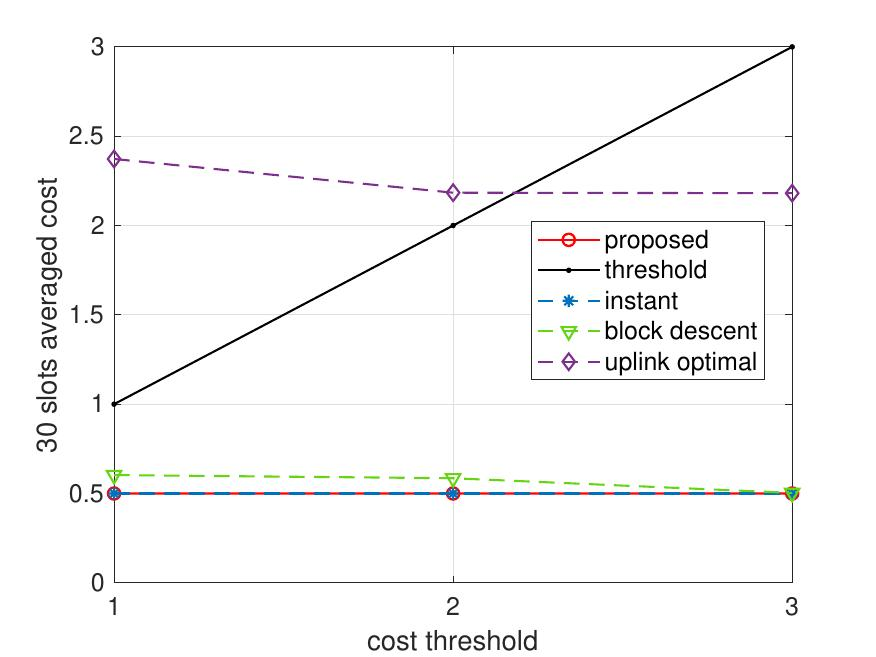}}
    \label{1a}
	  \subfloat[Time average delay with \protect\\different $B$]{
        \includegraphics[width=0.3\linewidth]{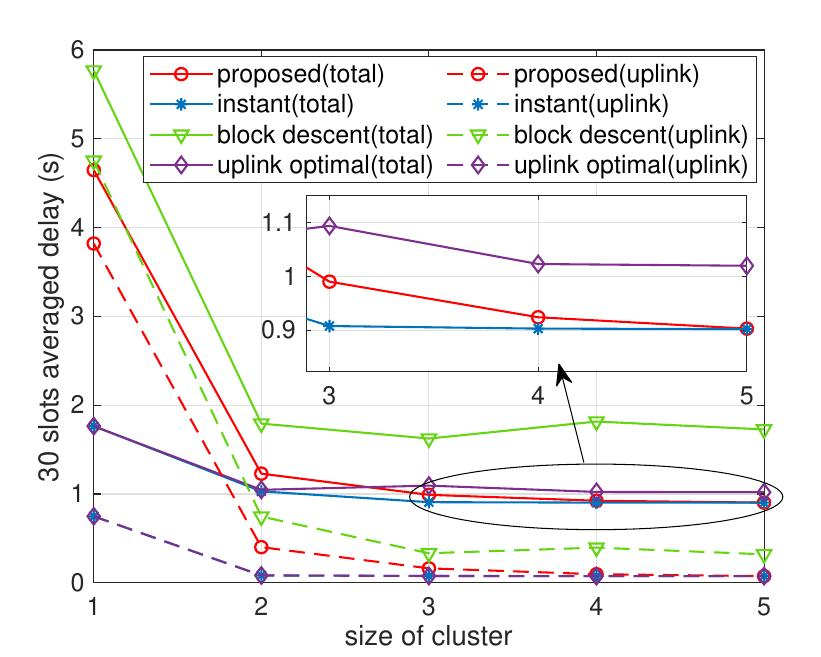}}
    \label{1b}
	  \subfloat[Time average cost \protect\\with different $B$]{
        \includegraphics[width=0.3\linewidth]{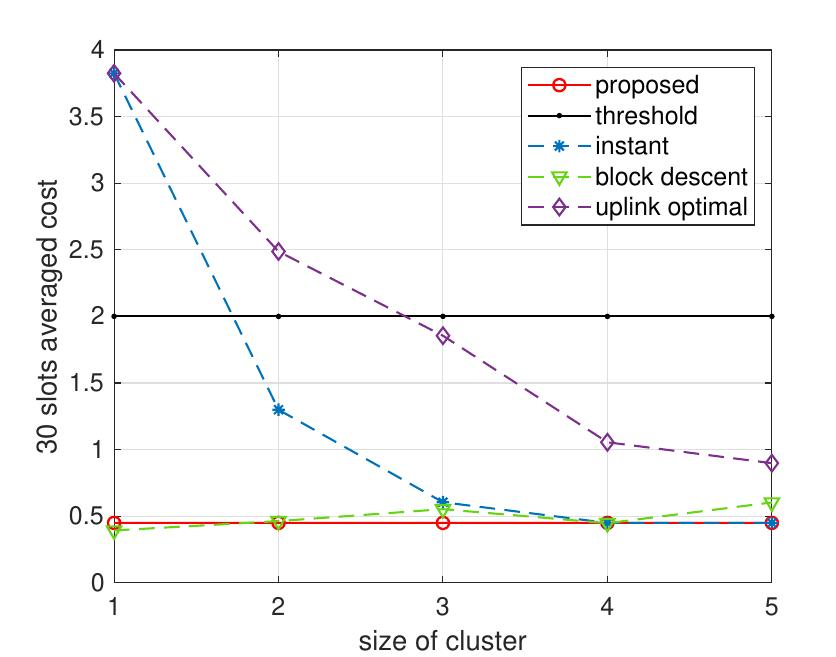}}
    \label{1c}

	  	  \caption{Performance comparison in the single-user scenario}
	  \label{fig1} 
	\end{figure*}

\subsection{Performance in the Single-User Scenario}

 \subsubsection{Long-Term Performance Analysis}
Next, to verify the effectiveness of the algorithm in the long-term process, we compare the long term delay and caching cost in 30 time slots for all four algorithms.

The long term average total delay and uplink delay of four algorithms are illustrated in Fig. 3(a). As time goes on, the average delay will gradually tend to a stable value. For the total delay, it can be seen that the total delay of the proposed algorithm can stay close to the optimal total delay, i.e., the total delay of instant. The total delay of block descent is the highest among all algorithms then followed by the total delay of uplink optimal. For the uplink delay, the curve of uplink optimal is coincide with the optimal curve, followed by the proposed algorithm. Although uplink optimal can obtain the best uplink delay performance, the overall delay performance is seriously affected due to excessive attention to the uplink process. The proposed algorithm can achieve the near-optimal total delay at the expense of acceptable uplink delay performance.

  Fig. 3(b) presents the long term average caching cost of four algorithms, where the caching cost threshold is set to $2$. It can be see that the average caching cost of uplink optimal is the highest among all algorithms, and far exceeds the caching cost threshold. This is because uplink optimal does not optimize the service caching strategy. It is worth noticing that our proposed algorithm can achieve the lowest caching cost, which can prove the necessity and optimality of the joint optimization.
  
.

\subsubsection{Impact of Caching Cost Thresholds}

Then we turn to explore the average delay and caching cost performance under different caching cost thresholds.

Fig. 3(c) shows the average total delay and  uplink delay under different caching cost thresholds. Each data point is the average value of 30 time slots. It can be seen that the total delay and the uplink of our proposed algorithm and the instant optimal can keep stable with the change of caching cost threshold. For the block descent and uplink optimal algorithm, there is also no obvious fluctuation in the delay with the increase of threshold. Therefore, it can be considered that the caching cost threshold has little effect on the delay performance.

In Fig. 3(d), we evaluate the average caching cost under different caching cost thresholds. Similar to Fig. 3(c), with the change of threshold, the average caching cost of all algorithms do not have a obvious fluctuation. There results prove that our proposed JO-CDSD algorithm can still remain the near-optimal 
average delay and caching cost performance in different caching cost threshold.

\subsubsection{Impact of BS Clustering Size}

Next, we further study the effect of different BS clustering size on the average delay and caching  cost.

Next, we compare the average total delay and uplink delay under different BS cluster sizes. It can be seen from Fig. 3(e) that when the BS cluster size increase, the average total delay, and uplink delay will decrease. This is because more BSs in the BS cluster can improve the performance of joint decoding, hence can improve the data transmission rate. On the other hand, more BSs will have more sufficient computing and caching resources to complete the offloading requests. Therefore, the increase in BS cluster size can improve the performance of data transmission and task processing. However, the trend of decreasing delay will gradually slow down, indicating that the delay gain brought by the BS cluster size is limited. This is because the cost of backhaul links bandwidth occupation will increases steadily with the BS cluster size. This result can give guidance for the selection of the optimal BS cluster size. In addition, when the BS cluster size increases, the delay of the proposed algorithm will be closer to the instant optimal algorithm. This is because the search space of the BS clustering becomes smaller, which is easier to converge and obtain the optimal clustering solution.

 Fig. 3(f) shows the average caching cost under different BS cluster sizes. It can be seen that when the BS cluster size increase, the average caching cost of our proposed algorithm and block descent can remain stable, and the average caching cost of instant optimal and uplink optimal will decrease rapidly. The results show that compared with traversal search based instant optimal method and uplink optimal, our proposed algorithm can always maintain the optimal caching cost stably.

\subsection{Performance in the Multi-User Scenario}
Moreover, we evaluate the effectiveness of our dichotomy JO-CDSD algorithm in the multi-user scenario. We assume that the offloading requests of three users correspond to the Zipf distribution of different parameters, which are Zipf(0.3,6), Zipf(0.2,6), and Zip(1,6), respectively. We compare the dichotomy-based JO-CDSD (\textbf{Algorithm 3}) with the dichotomy-based instant optimal and block descent algorithm.

\subsubsection{Long-Term Performance Analysis}
Fig. 4(a) shows the average total delay and uplink delay under corresponding time slots. It can be seen that the proposed algorithm can maintain the near-optimal delay performance, which is consistent with instant optimal, without obvious fluctuation due to the change in the number of request types. In contrast, block descent can not converge to the optimal value due to the increase of request types, resulting in a large fluctuation.

Fig. 4(b) presents the average caching cost in 30 time slots. Similar to Fig. 4(a), the average cost of the dichotomy-based JO-CDSD algorithm can maintain the near-optimal caching cost compared to block descent, which is far lower than the caching cost threshold. The results indicated that our algorithm is still effective in the multi-user scenario.

\begin{figure*}[tbp]
    \centering
    \captionsetup{justification=centering}
	  \subfloat[Time average delay]{
       \includegraphics[width=0.3\linewidth]{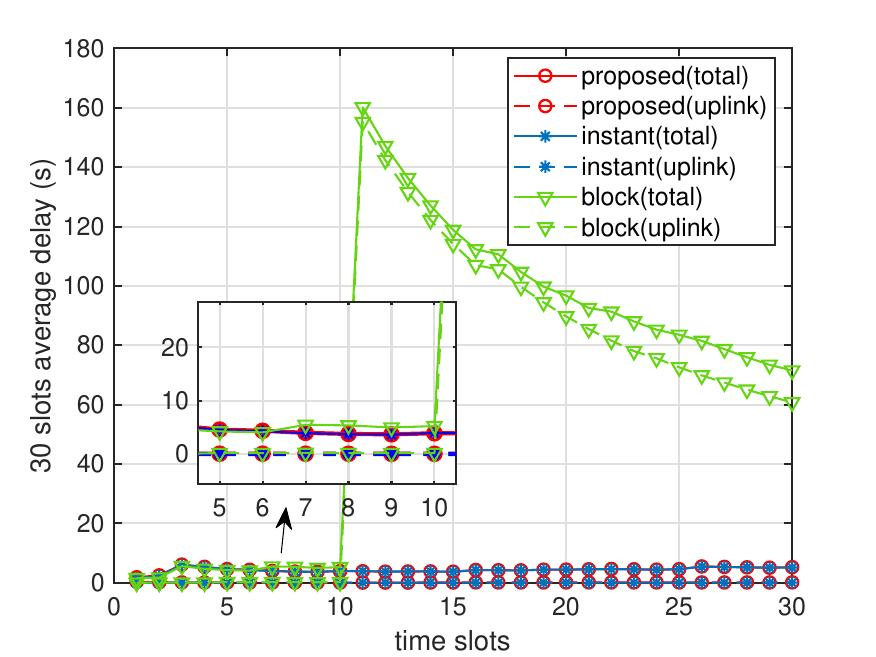}}
    \label{1a}
	  \subfloat[Time average cost]{
        \includegraphics[width=0.3\linewidth]{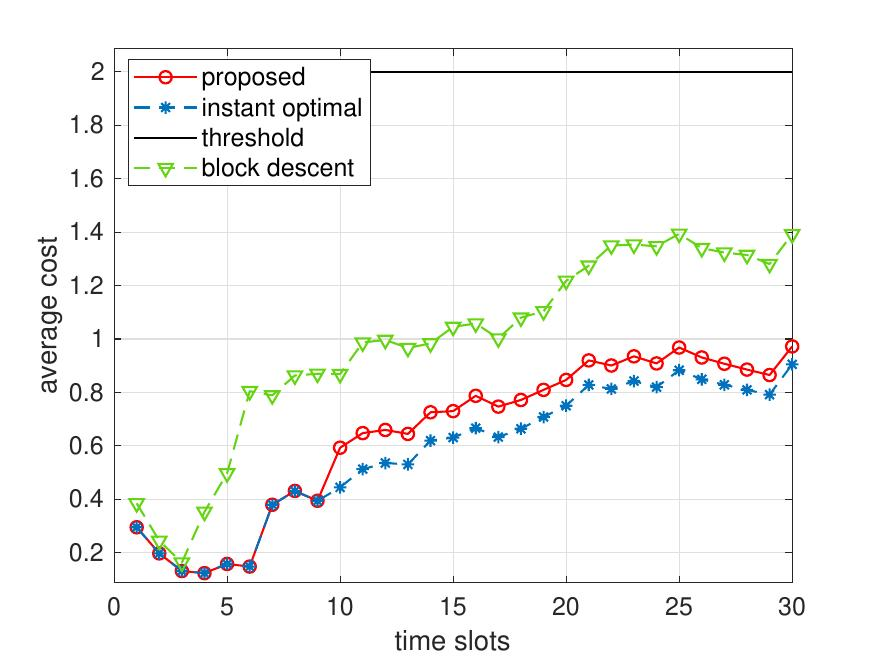}}
    \label{1b}
	  \subfloat[Time average delay with \protect\\different $cost^{th}$]{
        \includegraphics[width=0.3\linewidth]{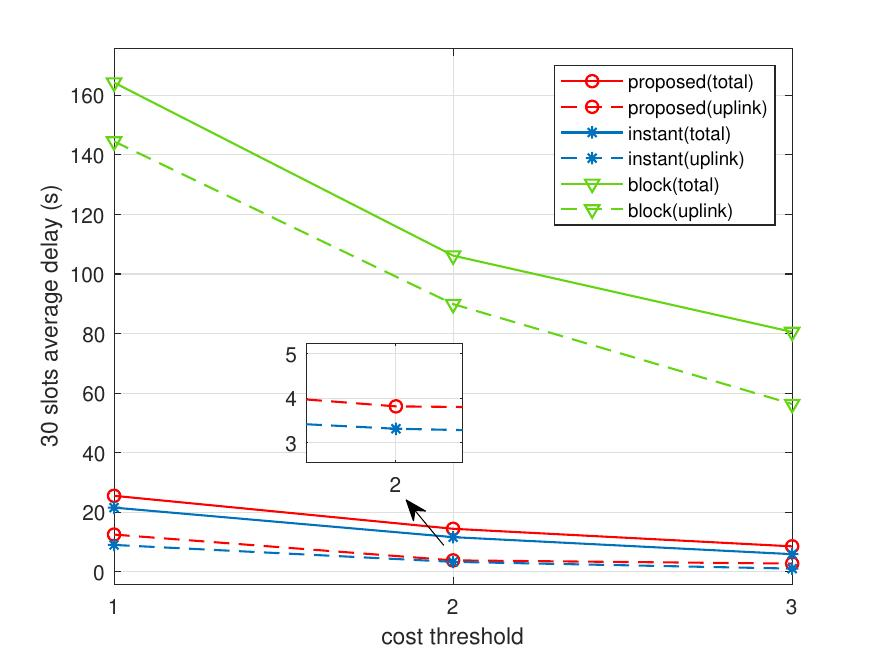}}
    \label{1c}
     \subfloat[Time average cost with \protect\\different $cost^{th}$]{
       \includegraphics[width=0.3\linewidth]{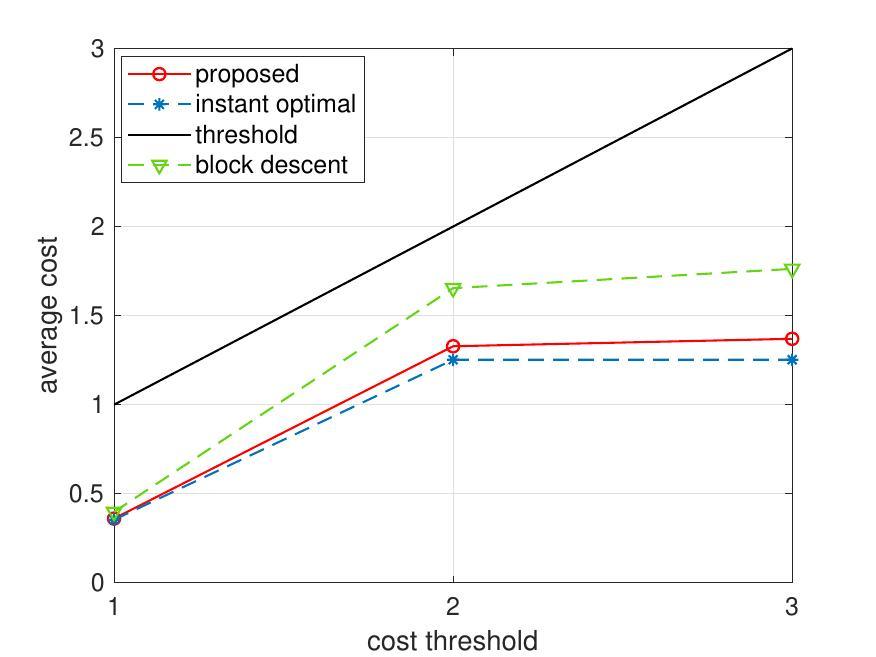}}
    \label{1a}
	  \subfloat[Time average delay with \protect\\different $B$]{
        \includegraphics[width=0.3\linewidth]{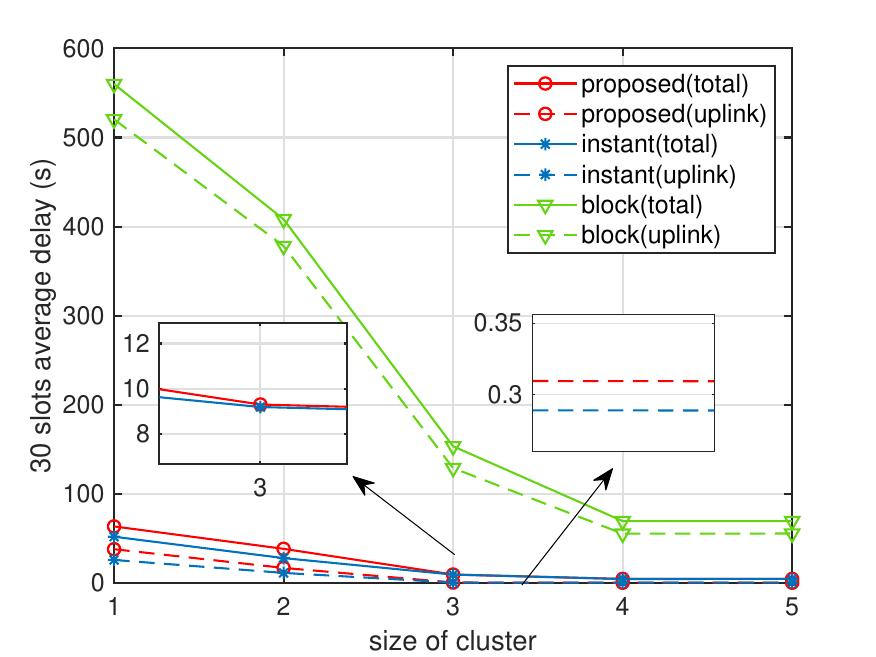}}
    \label{1b}
	  \subfloat[Time average cost \protect\\with different $B$]{
        \includegraphics[width=0.3\linewidth]{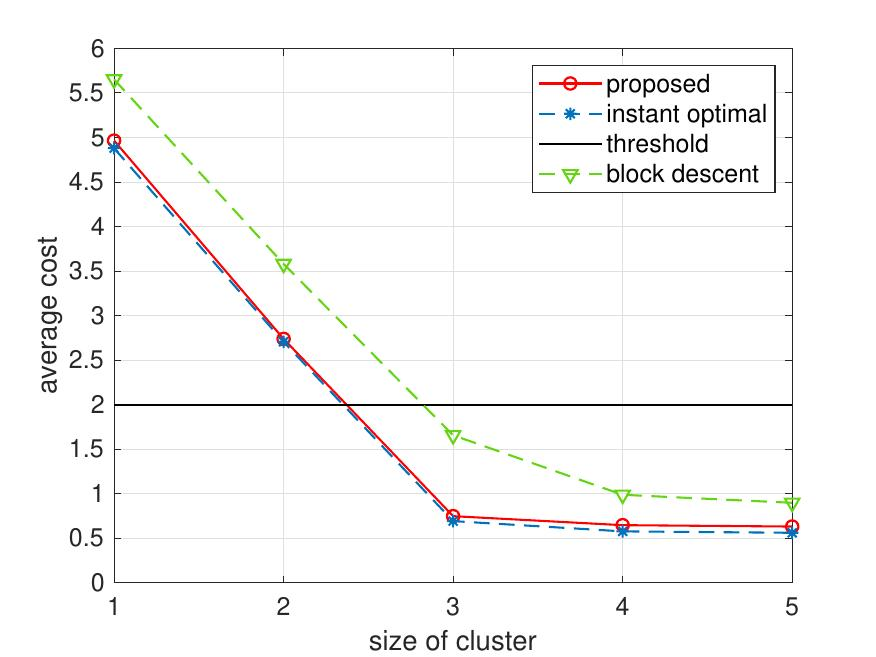}}
    \label{1c}

	  	  \caption{Performance comparison in the multi-user scenario}
	  \label{fig1} 
	\end{figure*}

\subsubsection{Impact of Caching Cost Thresholds}
Fig. 4(c) shows the average total delay and uplink delay under different caching cost thresholds. It can be seen that the caching cost threshold will not cause large performance fluctuation to the average delay under proposed and instant. On the contrary, when the threshold increases, the average delay in the block descent scenario will increase. It can be seen from the figure that our proposed algorithm can maintain the delay performance close to that of instant optimal.

Then Fig. 4(d) shows the average total delay and uplink delay under different caching cost thresholds. It can be seen that when the cost threshold increases, the average caching cost will increase. This is because the BS can find the optimal clustering strategy and caching strategy in a larger solution space under the premise of satisfying the threshold constraint. Similar to the trend in Fig.4 (c), the proposed algorithm can achieve the near-optimal caching cost, which is smaller than that of block descent.

It can be inferred that the change of the caching threshold will affect the average delay and cache cost in the multi-user scenario,  which is different from the single-user scenario. This is because multiple users will compete for the limited computing and caching resources of BSs, and the average caching cost will increase correspondingly.

\subsubsection{Impact of BS Clustering Size}
 In Fig. 4(e), we evaluate the average total delay and uplink delay under different BS cluster sizes. It can be seen that when the cluster size increases, the average uplink and total delay will decrease. Because a larger BS cluster will have more computing and caching resources, which can provide faster transmission and caching services for all users. Compared with block descent, the proposed algorithm can obtain near-optimal delay performance.

 The average total delay and uplink delay under different BS cluster sizes in shown in Fig. 4(f). It can be seen that when the cluster size increases, the average caching cost will decrease. When the cluster size reaches 3, the cost of all scenarios will be lower than the cost threshold. The proposed algorithm can still obtain the near-optimal caching cost performance. 
 
In summary, the proposed algorithm can get the near-optimal delay and cost performance at different parameters, which can prove the effectiveness of the proposed algorithm in the multi-user scenario.

\section{CONCLUSION And Future Work}
To overcome the system performance degradation brought by wireless transmission in the traditional cellular-based service caching networks, we propose a novel user-centric edge service caching framework to realize effective and reliable transmission and task processing for users. To make full advantage of user-centric MEC, a long-term average delay minimization problem is formulated by jointly optimizing the BS clustering and service caching decisions. Particularly, we transform the long-term optimization problem into multiple independent instantaneous optimization problems based on Lyapunov optimization. To solve the instantaneous problem, we decompose the problem into a primal problem and a master problem based on GBD framework. An online BS clustering and service caching algorithm (JO-CDSD) is proposed  to solve the primal and master problem alternatively. Simulation results show that th proposed algorithm can achieve significant average delay and caching cost gains compared with other benchmarks. Specifically, JO-CDSD can effectively reduce the long-term delay by at most $93.75\%$ and caching cost by at most $53.12\%$.

UCN and MEC can be a very interesting combination to provide an efficient, reliable, and low-cost transmission and computing service. As a new framework, there are still some open issues on long-term service caching problems in user-centric MEC. Firstly, offloading requests of users in multiple time slots are correlated in a short period. The correlation characteristics can be used to make short-term predictions for service requests, which is conducive to more accurate service caching. Secondly, when multiple users generate offloading requests at the same time, it will lead to contention for resources, making the coupling relationships of variables more complicated. Therefore, the service caching in UCN with multiple users needs to be further studied.

\bibliographystyle{IEEEtran}
\bibliography{ref}

\end{document}